\documentclass[aip,reprint,jcp,floatfix]{revtex4-1}

\usepackage{graphicx}
\usepackage{epstopdf}
\usepackage{mhchem}
\usepackage{amsmath}

\usepackage[usenames,dvipsnames]{color}

\begin{document}

\title{Ultraslow radiative cooling of C$_n^-$ ($n=3-5$)}

\author{James~N.~Bull}
\email{james.bull@uea.ac.uk}
\affiliation{School of Chemistry, Norwich Research Park, University of East Anglia, Norwich NR4 7TJ, United Kingdom}
\author{Michael~S.~Scholz}
\affiliation{School of Chemistry, University of Melbourne, Parkville, VIC 3010, Australia}
\author{Eduardo~Carrascosa}
\affiliation{Laboratoire de Chimie Physique Mol\'{e}culaire, \'{E}cole Polytechnique F\'{e}d\'{e}rale de Lausanne, EPFL SB ISIC LCPM, Station 6, CH-1015 Lausanne, Switzerland}
\author{Moa~K.~Kristiansson}
\affiliation{Department of Physics, Stockholm University, SE-10691 Stockholm, Sweden}
\author{Gustav~Eklund}
\affiliation{Department of Physics, Stockholm University, SE-10691 Stockholm, Sweden}
\author{Najeeb~Punnakayathil}
\affiliation{Department of Physics, Stockholm University, SE-10691 Stockholm, Sweden}
\author{Nathalie~de~Ruette}
\affiliation{Department of Physics, Stockholm University, SE-10691 Stockholm, Sweden}
\author{Henning~Zettergren}
\affiliation{Department of Physics, Stockholm University, SE-10691 Stockholm, Sweden}
\author{Henning~T.~Schmidt}
\affiliation{Department of Physics, Stockholm University, SE-10691 Stockholm, Sweden}
\author{Henrik~Cederquist}
\affiliation{Department of Physics, Stockholm University, SE-10691 Stockholm, Sweden}
\author{Mark~H.~Stockett}
\email{mark.stockett@fysik.su.se}
\affiliation{Department of Physics, Stockholm University, SE-10691 Stockholm, Sweden}
\begin{abstract}
Ultraslow radiative cooling lifetimes and adiabatic detachment energies for three astrochemically relevant anions, C$_n^-$ ($n=3-5$), are measured using the Double ElectroStatic Ion Ring ExpEriment (DESIREE) infrastructure at Stockholm University. DESIREE maintains a background pressure of $\approx$10$^{-14}$\,mbar and temperature of $\approx$13\,K, allowing storage of mass-selected ions for hours and providing conditions coined a ``molecular cloud in a box". Here, we construct two-dimensional (2D) photodetachment spectra for the target anions by recording photodetachment signal as a function of irradiation wavelength and ion storage time (seconds to minute timescale). Ion cooling lifetimes, which are associated with infrared radiative emission, are extracted from the 2D photodetachment spectrum for each ion by tracking the disappearance of vibrational hot-band signal with ion storage time, giving $\frac{1}{e}$ cooling lifetimes of 3.1$\pm$0.1\,s (C$_3^-$), 6.8$\pm$0.5\,s (C$_4^-$) and 24$\pm$5\,s (C$_5^-$). Fits of the photodetachment spectra for cold ions, i.e. those stored for at least 30\,s, provides adiabatic detachment energies in good agreement with values from laser photoelectron spectroscopy. Ion cooling lifetimes are simulated using a Simple Harmonic Cascade model, finding good agreement with experiment and providing a mode-by-mode understanding of the radiative cooling properties. The 2D photodetachment strategy and radiative cooling modeling developed in this study could be applied to investigate the ultraslow cooling dynamics of wide range of molecular anions.
\end{abstract}

\maketitle
\section{Introduction}
Which molecular anions exist in space? What are their formation mechanisms and life cycles? These are two long-standing questions in astrochemistry.\cite{Larson:2012,Millar:2017} Prior to a decade and a half ago, H$^{-}$ was the only anion thought to play a prominent role in the interstellar medium (ISM). In 2006, the first molecular anion, C$_{6}$H$^{-}$, was discovered by comparing astronomical line spectra with gas-phase action spectra recorded in the laboratory.\cite{C6m} Over the next four years there were five further identifications: \ce{C4H-},\cite{C4hm} \ce{C8H-},\cite{C8hm} \ce{CN-},\cite{CNm} \ce{C3N-},\cite{C3Nm} and \ce{C5N-}.\cite{C5Nm} Vibrationally excited \ce{C6H-} was also detected alongside \ce{C5N-}.\cite{C5Nm} However, despite increasing interest in the role of molecular anions in space, there has been a stall in new identifications. It is thought that the discovery of new molecular anions is thwarted by a lack of understanding of the formation mechanism(s) and dynamical properties of both the anions known to exist in the ISM and new anions yet to be assigned.\cite{Millar:2017,herbst:2014} Dynamical properties in this context include electron capture cross-sections, electronic internal conversion efficiencies and couplings between dipole-bound and valence-localised states, cross-sections for neutralization reactions with cations, and the rates of radiative cooling.\cite{Carelli:2013,Roueff:2013} As an example of the need for reliable measurements of the dynamical properties of astrochemically relevant anions, in a discussion on radiative electron attachment (which involves formation of a vibrationally excited ground state ion that must cool) Herbst\cite{Herbst:2009} remarked ``\emph{The discovery of molecular anions [in space] has generated the need to  include  their formation and destruction in chemical models ... the larger [carbonaceous] molecular anions detected ($n$ = 6, 8) have higher abundances relative to their neutral precursors because the radiative attachment rate increases with the number of degrees of freedom of the anion. However, their rate estimates are quite uncertain and experimental studies are highly welcome.}" Although gas-phase action spectroscopies can provide data on electronic transitions and detachment energies for carbonaceous anions,\cite{Gerlich:2006} their dynamical properties such as infrared (IR) radiative cooling lifetimes are more difficult to measure because hot anions need to be isolated (i.e. free from collision) for periods of milliseconds to minutes. These conditions are not attainable using conventional ion traps.


Here, we used the Double ElectroStatic Ion Ring ExpEriment (DESIREE) infrastructure at Stockholm University to characterize the radiative cooling lifetimes and adiabatic detachment energies (ADEs) of C$_n^-$ ($n=3-5$) by monitoring the intensity of vibrational hot bands near the electron detachment threshold with ion storage time. The present investigation targeted the C$_n^-$ ($n=3-5$) species because they are likely ISM anions. In particular, they possess similar bonding and electronic structure to the molecular anions already known to exist in the ISM, they could be formed through either photodissociation or dissociative electron attachment mechanisms,\cite{Larson:2012,Millar:2017} and neutral C$_{3}$ and C$_{5}$ are known interstellar molecules.\cite{Hinkle:1988,Bernath:1989} Although anions are unlikely to be significant astrochemical species in `photon-dominated regions' (PDRs, e.g. diffuse clouds) due to facile destruction by photodetachment with visible and ultraviolet light,\cite{Millar:2017,Khamesian:2016} the abundance of anions in dark clouds (e.g. C$_{6}^{-}$ in L1527)\cite{Sakai:2007} has been shown to reach nearly 10\% of that for the corresponding neutral molecule, suggesting that negative charge in photon-free regions of space is more likely in the form of anions than free electrons. Cold dark molecular clouds have temperatures of 10--20\,K.\cite{Bergin:2007} The normal operating temperature of DESIREE ($\approx$13\,K) is squarely within this range. \cite{Schmidt2017}

In the absence of collisional quenching, the spontaneous cooling dynamics of hot molecular anions such as C$_n^-$ ($n=3-5$) can be divided into three time ($t$) regimes:
\begin{itemize}
\item Statistical regime I ($t\leq$10$^{-3}$\,s): Internal energy is high, e.g. several electron-volts above the thresholds for dissociation, thermionic emission,\cite{Zhao:1996} and recurrent/Poincar\'{e} fluorescence.\cite{Leger:1988} The energy threshold for thermionic emission is the ADE, which for C$_n^-$ ($n=3-5$) is $\approx$2--4\,eV, and the energy threshold for dissociation can be approximated by the lowest bond dissociation energy ($\approx$3--4\,eV). Both thermionic emission and dissociation lead to destruction of the anion. Recurrent fluorescence, which is known for C$_4^-$ and C$_6^-$,\cite{Ito:2014,Kono2015,Ebara2016,Kono:2018} involves inverse internal conversion to an electronic excited state (situated below the ADE, e.g. $\approx$2--3\,eV) followed by radiative emission.\cite{Shiromaru:2015} 

\item Slow regime II ($t\approx$10$^{-3}$--1\,s): Internal energy is in the vicinity of the lowest thresholds for the cooling mechanisms important for Regime I, e.g. $\sim$2\,eV. In addition to slow cooling through these mechanisms, radiative emission due to vibrational transitions becomes important.

\item Ultraslow regime III ($t\gtrsim$1\,s): Internal energy is below the thresholds for dissociation, thermionic emission and recurrent fluorescence. Ions cool only through radiative vibrational (IR) transitions and rotational (microwave) transitions, with vibrational cooling occurring much faster than rotational cooling.\cite{Schmidt2017,Meyer2017} Cooling dynamics in this regime have been explored for only a few small anions due to technical challenges associated with isolating ions for durations extending to minutes and maintaining low background temperatures.\cite{vh2016,RICE2017,Chartkunchand2016,Connor2016,Hansen2017,Anderson:2018}
\end{itemize}

While the target anions have been intensively studied by several groups in recent years, all studies used room-temperature electrostatic ion storage rings or beam traps and were limited to characterizing cooling dynamics occurring on sub-second timescales.\cite{Naaman2000,Goto2013,Chandrasekaran2014,Kono2015,Saha2017,Kono:2018} In the present study, we have used the DESIREE infrastructure to investigate the cooling dynamics of the target anions on the ultraslow, $t\,\gtrsim$1\,s timescale. Our strategy involved storing ions for up to $\approx$1\,minute and using photodetachment spectroscopy to monitor the intensity of hot bands with ion storage time, providing an indirect characterization of cooling lifetimes. The $\frac{1}{e}$ ion cooling lifetimes, which are attributed to IR radiative emission, are well-described by a simple harmonic cascade model of this process. Fits of the cold photodetachment spectra associated with ions stored for at least 30\,s to the Wigner threshold law demonstrate an alternative, cryogenic method for obtaining ADE values.

\begin{figure}[!b]
\centering
  \includegraphics[width=0.99\columnwidth]{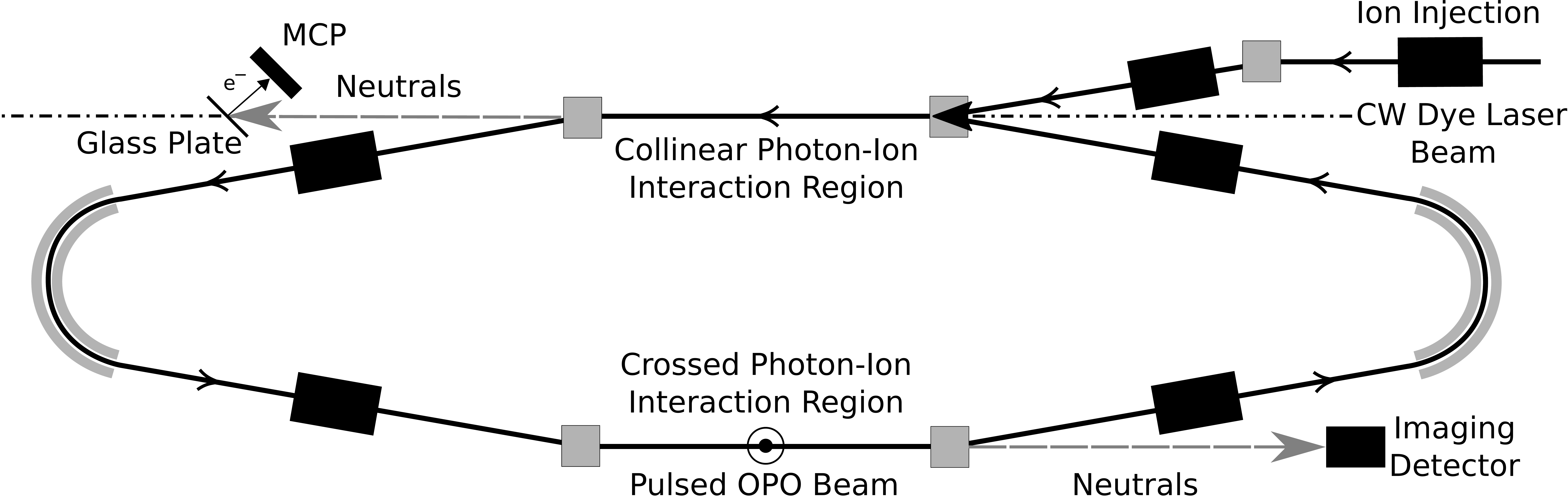}
  \caption{The symmetric ion storage ring in DESIREE. Neutral particles formed by photodetachment in the lower straight region (crossed-beam geometry with the optical parametric oscillator, OPO) are detected in their forward directions with the `Imaging Detector'. In another set of measurements on C$_3^-$, light from a cw dye laser was merged colinearly with the stored ion beam in the upper straight section. The wavelength of the cw light was tuned to be in resonance with the vibrational hot band of C$_{3}^{-}$ (620.5\,nm). The ring circumference is 8.6\,m and each straight section has a length of 0.96\,m.}
  \label{fig:desiree}
\end{figure}

\section{Methods}
DESIREE is a cryogenic dual electrostatic ion storage ring facility located at the Department of Physics, Stockholm University.\cite{Thomas2011,Schmidt:2013} The major components constituting the so-called `symmetric' storage ring are schematically illustrated in FIG.\,\ref{fig:desiree}. The interior of the ring is cooled to $\approx$13~K by compressed helium refrigerators and is isolated from external thermal radiation by several layers of insulation.\cite{Schmidt2017} Vacuum is maintained at a background pressure of $\approx$10$^{-14}$\,mbar using cryopumping combined with turbomolecular pumps and oil-free backing pumps. These ultrahigh vacuum conditions allow storage of keV ion beams for hours.\cite{Backstrom2015}  In the present experiments, the target anions [C$_n^-$ ($n=3-5$)] were produced using a caesium sputtering ion source with a graphite cathode.\cite{Schmidt:2013} This process generates ions with a high degree of rovibrational excitation, i.e. source-heated ions. The nascent ions were accelerated to 10~keV, selected according to their mass-to-charge ratio using a bending magnet, and injected into the symmetric ion storage ring. Transport from the source to the ring takes $\approx$100\,$\mu$s. The $\frac{1}{e}$ beam storage lifetimes were measured at 540$\pm$30\,s for C$_3^-$ and 570$\pm$30\,s for C$_5^-$ (see Supporting Information). Although the beam storage lifetime for C$_4^-$ was not measured in this study, we expect a similar lifetime to those for C$_3^-$ and C$_5^-$. These beam storage lifetimes are limited by loss of ions through collisions with background gas.\cite{Schmidt:2013} 

\subsection{One-color experiments}
In the one-color experiments, stored ions were irradiated with tunable-wavelength light from an optical parametric oscillator (OPO, EKSPLA NT342B, 10\,Hz) using a crossed-beam geometry through one of the straight sections of the ion storage ring (see FIG.\,1). Any neutral particles formed through photodetachment or photodissociation are unaffected by the ring's electrostatic steering fields and impact on a micro-channel plate (MCP) detector (`Imaging Detector' in FIG.\,\ref{fig:desiree}).\cite{Rosen:2007} Signal from the MCP detector was gated using a 1~$\mu$s duration pulse that was slightly delayed with respect to the OPO pulse to account for the neutral particle's flight time from the interaction region to the detector. The purpose of the gate was to eliminate signal from scattered OPO light striking the detector and to minimize background counts from collision-induced detachment events due to the residual gas consisting of $\sim$10$^4$ H$_{2}$ molecules per cm$^{3}$. The OPO wavelengths were calibrated using an optical spectrograph (Avantes AvaSpec-3648), which was itself calibrated against a wavemeter (HighFinesse WS-8) via a diode laser (632.6~nm). The irradiation wavelength was stepped in 0.5\,nm increments (2\,nm for C$_4^-$) between ion injections for a given ion storage time, providing a two-dimensional (2D) photodetachment spectrum, i.e. a series of photodetachment spectra as a function of wavelength and ion storage time (see Refs\,\citenum{Schmidt2017,Meyer2017} for a similar procedure applied to rotational cooling of OH$^{-}$). For a given ion, the time evolution of the photodetachment yield at a specific wavelength or range of wavelengths can be obtained by taking a wavelength slice through the 2D photodetachment spectrum.

Part of our interpretation applied Principal Component Analysis (PCA)\cite{pca:2002} to the 2D photodetachment spectrum for each ion. PCA is a common statistical procedure that decomposes a multi-demensional data set $X$ into a set of orthogonal principal components (PCs) which are the eigenvectors of the covariance matrix $X^TX$. The eigenvalues associated with each PC relate to the fraction of the variation in $X$ that is explained by each PC and the principal values (PVs, the projection of $X$ on its PCs) give the wieght of each PC as a function of time. In the present case, the PCs may be thought of as the underlying spectra that describe the evolution of the photodetachment spectra with ion storage time, with a time invariant background due to photodetachment signal from cold ions (or nearly time invariant because of a finite ion beam storage lifetime).\cite{Stockett2019} The cooling lifetimes obtained from the PVs should be considered wavelength-averaged values since each probe wavelength provides slightly different ion cooling lifetime due to a distribution of internal vibrational energies in the stored ion beam.

\subsection{Deplete-probe experiments with C$_3^-$}
Deplete-probe experiments on source-heated C$_3^-$ were performed by adapting the procedure recently described by Schmidt\,\textit{et al.}\cite{Schmidt2017}, where the effect of the depletion laser was to preferentially photodetach rotationally excited ions and thus reduce the measured ion cooling lifetimes. Depletion involved intercepting the stored ion beam with 620.5\,nm light from a cw laser (Coherent 899 ring dye laser) using a merged-beam geometry in the straight section of the ion storage ring opposite the OPO light interaction region (see FIG.\,\ref{fig:desiree}). The depletion laser wavelength (620.5\,nm or 1.998\,eV) was chosen to be close to the ADE from the present measurements (see below) because the photodetachment cross-section for vibrationally excited (hot band) C$_3^-$ is much larger than that for cold ions.

\subsection{Adiabatic detachment energies}
The adiabatic detachment energy (ADE) for each anion was extracted from cold photodetachment spectrum assuming fit with the Wigner threshold law:\cite{Wigner:1948,Farley:1989}
\begin{equation}
\sigma_{PD} = (KE)^{L + \frac{1}{2}},
\end{equation}

\noindent where $\sigma_{PD}$ is the photodetachment cross-section, $KE$ is the kinetic energy of the ejected electron (energy in excess of the ADE for a direct photodetachment process), and $L$ is the angular momentum of the outgoing electron. For the present systems which involves photodetachment from $\pi$ molecular orbital, we find that $L=2$ ($d$ wave photoelectron) provides best fit to the experimental data. The ADE is taken to be the energy at which the Wigner threshold law fit exceeds 3$\sigma$ of the baseline signal. We note the above expression is strictly valid for atomic species; best fit values of $L$ can deviate from integers for molecules -- see example fit for C$_{3}^{-}$ in the Supporting Information. 

\subsection{Radiative cooling lifetime modeling}
Spontaneous cooling in the present experiments is presumed to occur through IR radiative emission. A simple harmonic cascade (SHC) model was developed to interpret the experimental results. The model assumes vibrational density of states $\rho$ computed using the Beyer-Swinehart algorithm using anharmonic or scaled harmonic vibrational mode frequencies $\nu_{s}$ calculated at the $\omega$B97X-D//aug-cc-pVTZ level of theory with Q-Chem\,4.4 (see Supporting Information).\cite{wb97xd,acct,qchem} For a given mode $s$, the IR radiative cooling rate coefficient, assuming only transitions where $\Delta v_{s}=-1$ are allowed, with $v$ being the vibrational quantum number, is \cite{Chandrasekaran2014}

\begin{equation}
k_s(E)=A_s^{10}\sum_{v=1}^{v\leq E/h\nu_s} \frac{\rho(E-vh\nu_s)}{\rho(E)},
\label{eq_k}
\end{equation}

\noindent where $E$ is the energy of a given vibrational state, $h$ is Plank's constant, and the summation is over $v$ ($v=0$ and $1$ are the ground and first excited vibrational states of mode $s$, respectively). The Einstein coefficients $A_s^{10}$ were calculated at the $\omega$B97X-D//aug-cc-pVTZ within the harmonic approximation (see Supporting Information). Starting from an initial Boltzmann distribution of vibrational energy $g(E,t=0)$ corresponding to 1000~K, the population in each level was recalculated at each simulation timestep. The model allowed for two treatments of intramolecular vibrational energy redistribution (IVR),\cite{Nesbitt:1996} i.e. statistical randomization of vibrational energy with time, $t$:\\\\
(i) IVR is negligible or slow compared with radiative cooling -- the population of each mode is explicitly tracked according to the expression below:

\begin{equation}
\begin{aligned}
g(E,t+dt) = \sum_s g(E,t)e^{-k_s(E)dt}\\ + \sum_s g(E+h\nu_s,t)(1-e^{-k_s(E+h\nu_s)dt});
\end{aligned}
\end{equation}
(ii) IVR is fast compared with radiative cooling -- vibrational energy is statistically redistributed each simulation time step and the total energy emitted radiatively at each time step is:\\
\begin{equation}
dE_{tot}/dt=-\int g(E,t)\sum_s h\nu_sk_s(E)dE,
\end{equation}

\noindent where the total energy remaining in the ensemble as a function of time $E_{tot}(t)=\int Eg(E,t)dE$ was taken as an indicator of the progress of cooling. Given that the vibrational energy quanta are small and the number of stored ions is large, level occupation numbers were treated as continuous quantities. We expect that case (ii) should be most relevant for the present source-heated anions because ion cooling lifetimes are long (seconds timescale) compared with the expected timescale for IVR (nanoseconds to millisecond timescale). 

The SHC modeling starts from a hot ensemble and simulates the internal energy as a function of ion storage time. For case (i), the internal vibrational energy reached a non-zero asymptotic value because any population that was portioned to IR inactive modes is not emitted radiatively. For case (ii), because the lowest frequency vibrational modes for each anion are IR active, all vibrational energy in excess of the zero-point energy can be liberated and thus the model goes asymptotically to zero vibrational energy at long times. To compare results from the SHC model with experiment for which there is non-zero photodetachment signal at long ion storage times for wavelengths shorter than the ADE due to photodetachment from cold ions, it was necessary to add an asymptote offset equal to the value extracted from an exponential fit of the experimental data. Furthermore, it was found that the initial temperature assumed in the SHC model (e.g. 500\,--\,5000\,K) altered the cooling dynamics only on timescales much faster than those probed in the present experiments, e.g. milliseconds.

It is worth noting that we found use of the commonly cited harmonic frequencies and intensities from Szczepanski \emph{et al.}\cite{Szczepanski:1997} calculated at the B3LYP/6-31G* level of theory within the SHC framework produced qualitatively similar results to that presented in this study, but required scaling the $A^{10}_{s}$ coefficients with factor 0.5 to best align modeled ion cooling lifetimes with experimental values.

\section{Results and discussion}
\subsection{Tricarbon anion, C$_3^-$}

\begin{figure}[!t]
\centering
  \includegraphics[width=0.99\columnwidth]{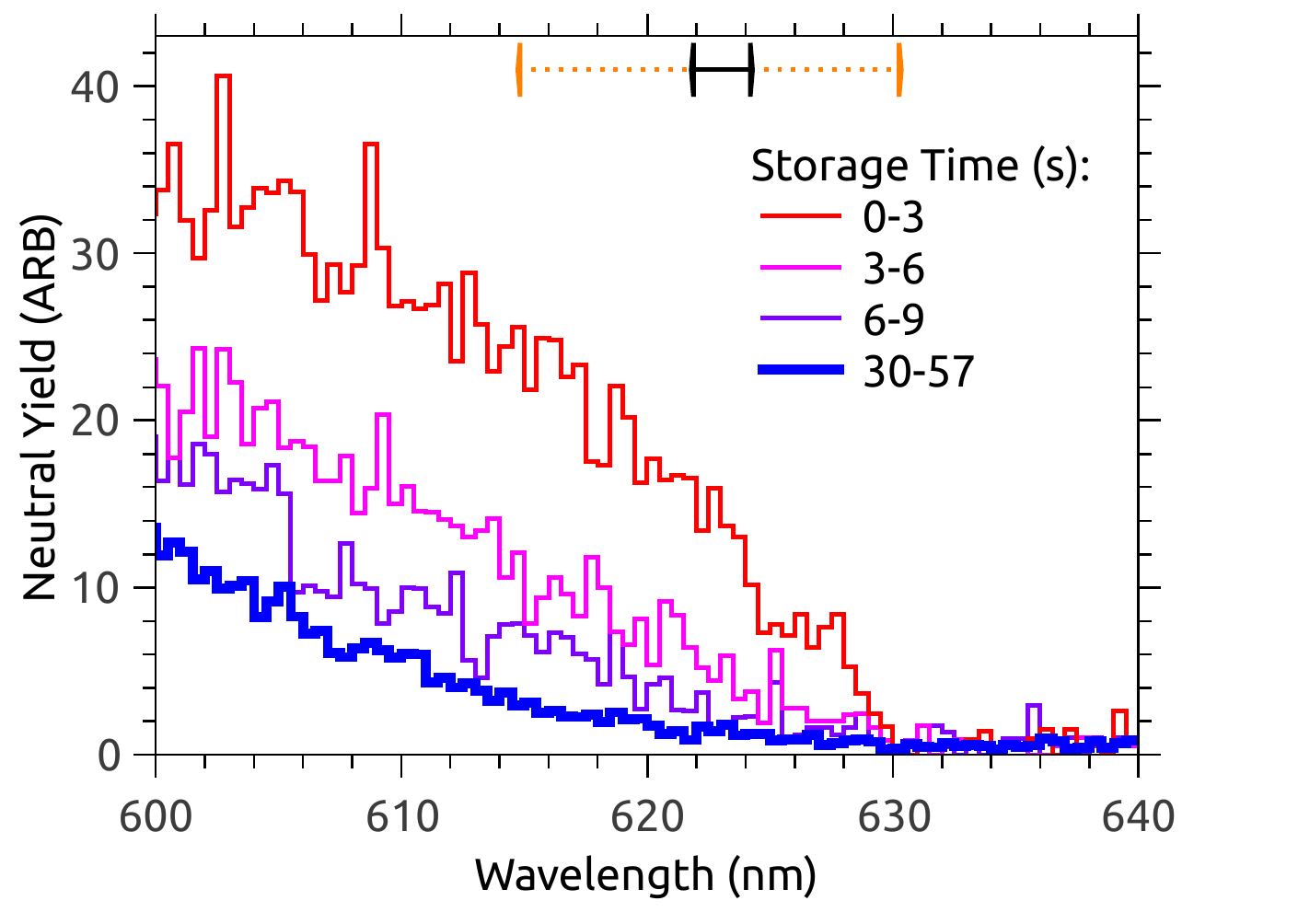}
  \includegraphics[width=0.99\columnwidth]{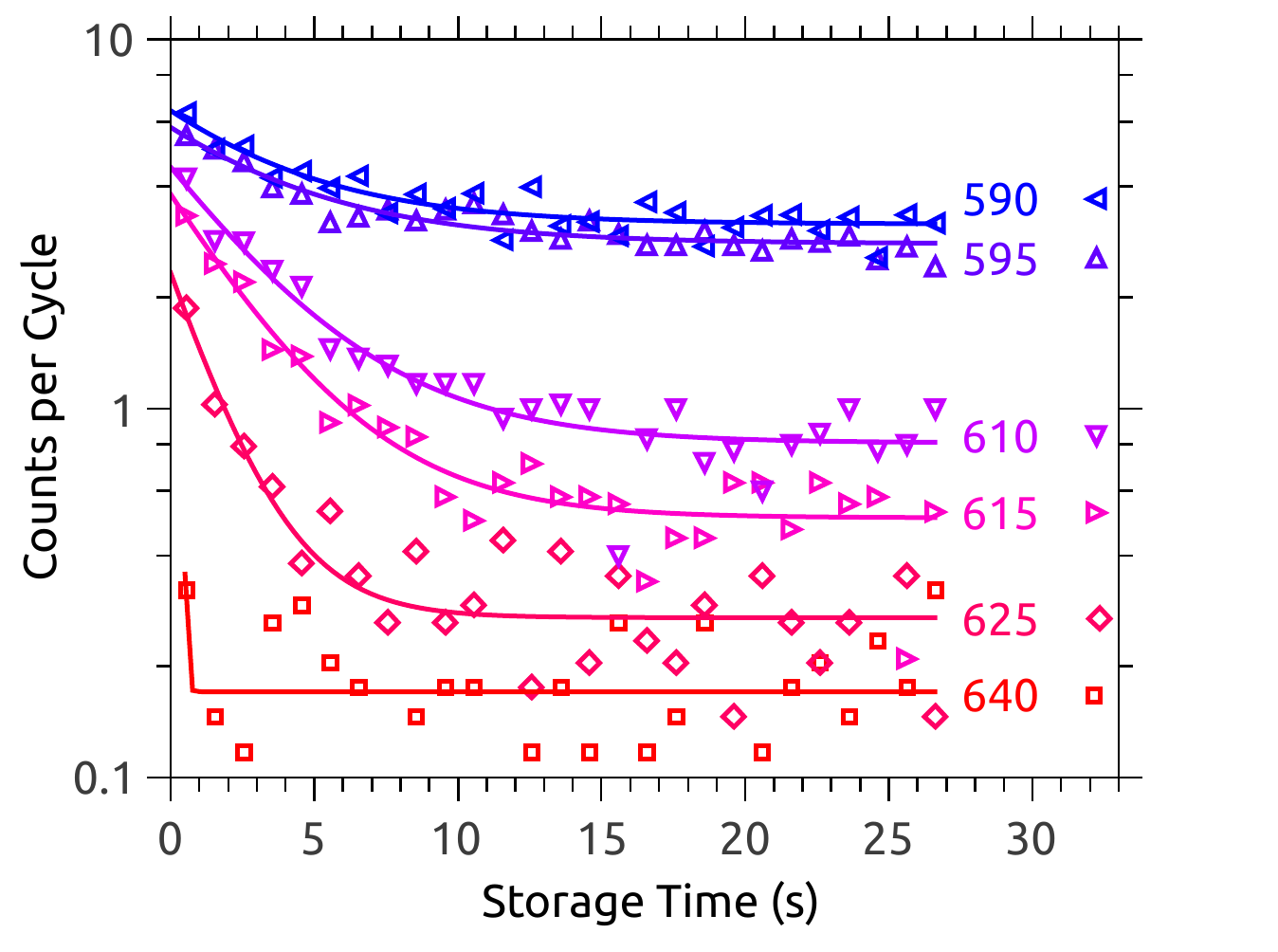}
  \caption{Upper: Time-binned photodetachment spectra for C$_3^-$, recorded by monitoring the yield of neutral particles with wavelength of light. The black bar represents the ADE and uncertainty determined for ions stored at least 30~s, and the orange bar corresponds to the photoelectron spectroscopy value from Ref.\citenum{Arnold:1991}. Lower: Decay of photodetachment signal with ion storage time at selected probe wavelengths (note the log scale). Time constants for single-exponential fits in the lower panel are 4.3$\pm$0.8\,s (590\,nm), 4.6$\pm$0.7\,s (595\,nm), 3.8$\pm$0.4\,s (610\,nm), 3.2$\pm$0.2\,s (615\,nm) and 1.8$\pm$0.2\,s (625\,nm).}
  \label{fig_c3raw}
\end{figure}

Photodetachment spectra for C$_3^-$ as a function of ion storage time are shown in FIG.\,\ref{fig_c3raw}, upper. The spectra have been divided into four time bins, with the 0--3\,s time bin corresponding to ions recently injected into the storage ring and the 30--57\,s time bin corresponding to ions that have been stored for at least 30\,s. The complete 2D photodetachment spectrum is shown in the Supporting Information. The time-binned photodetachment spectra show a broad feature over the expected ADE (1.99$\pm$0.025\,eV or 622$\pm$8\,nm from photoelectron spectroscopy)\cite{Arnold:1991} due to vibrationally excited C$_3^-$ ions that cool over the first $<$30\,s, providing a `cold' photodetachment spectrum (30--57\,s spectrum). A fit of the cold photodetachment spectrum with the Wigner threshold law gave ADE = 1.990$\pm$0.005\,eV (623$\pm$0.6\,nm), which is within error of the earlier photoelectron spectroscopy determination.

The lower panel of FIG.\,\ref{fig_c3raw} shows photodetachment signal with ion storage time (1\,s time bins) at six selected wavelengths. Fits with a single-exponential decay curve gave lifetimes ranging from 4.3$\pm$0.8~s (590~nm) to 1.8$\pm$0.2~s (615~nm), demonstrating that cooling occurs more slowly for anions probed at shorter wavelength, i.e. those closer to the detachment threshold. This is readily interpreted in terms of EQN\,\ref{eq_k}, which indicates that the cooling rate rapidly increases with vibrational excitation and the variation in ion cooling lifetime is due to the stored ion beam having a distribution of internal vibrational energies.

\begin{figure}[!t]
\centering
  \includegraphics[width=0.99\columnwidth]{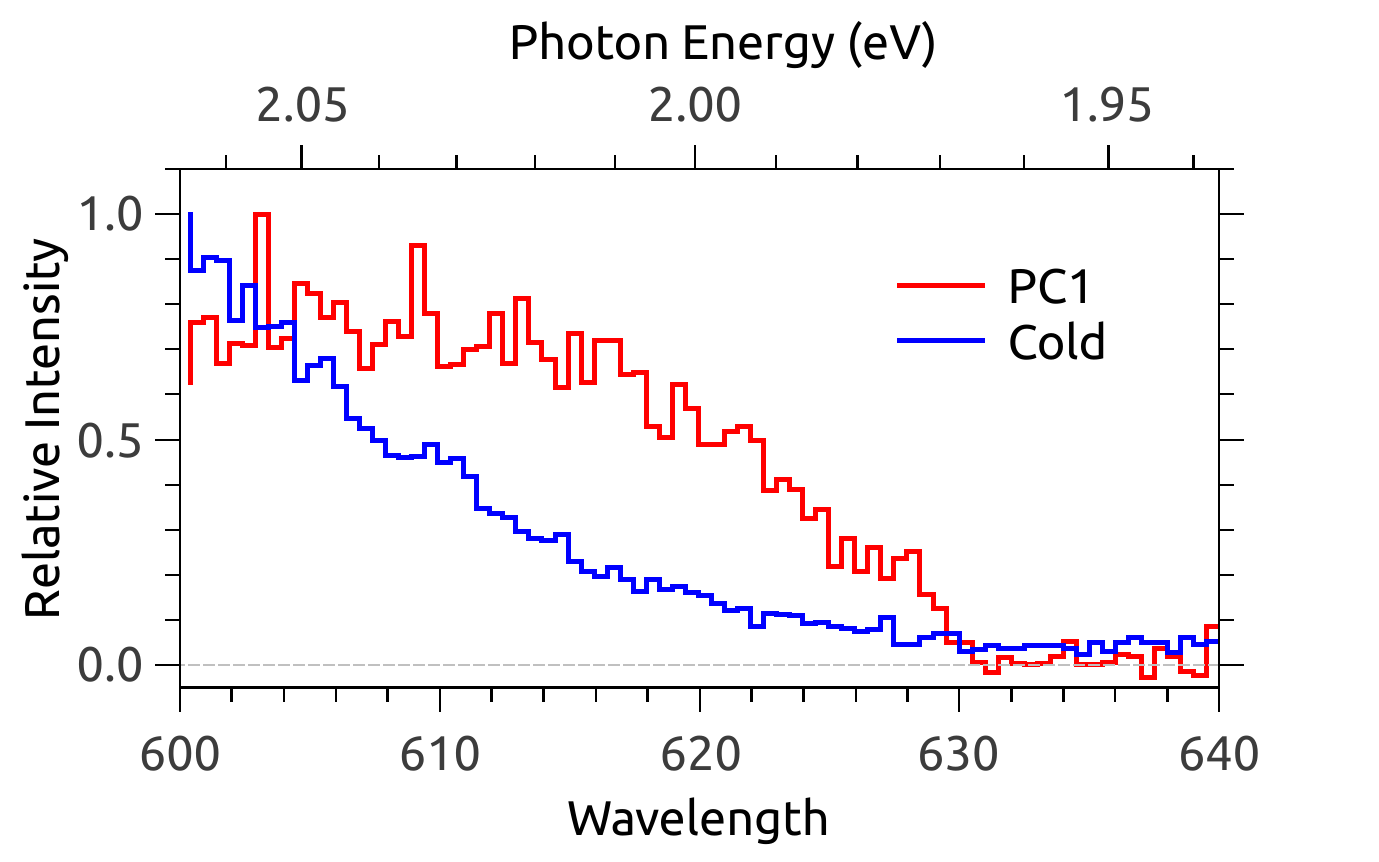}
  \includegraphics[width=0.99\columnwidth]{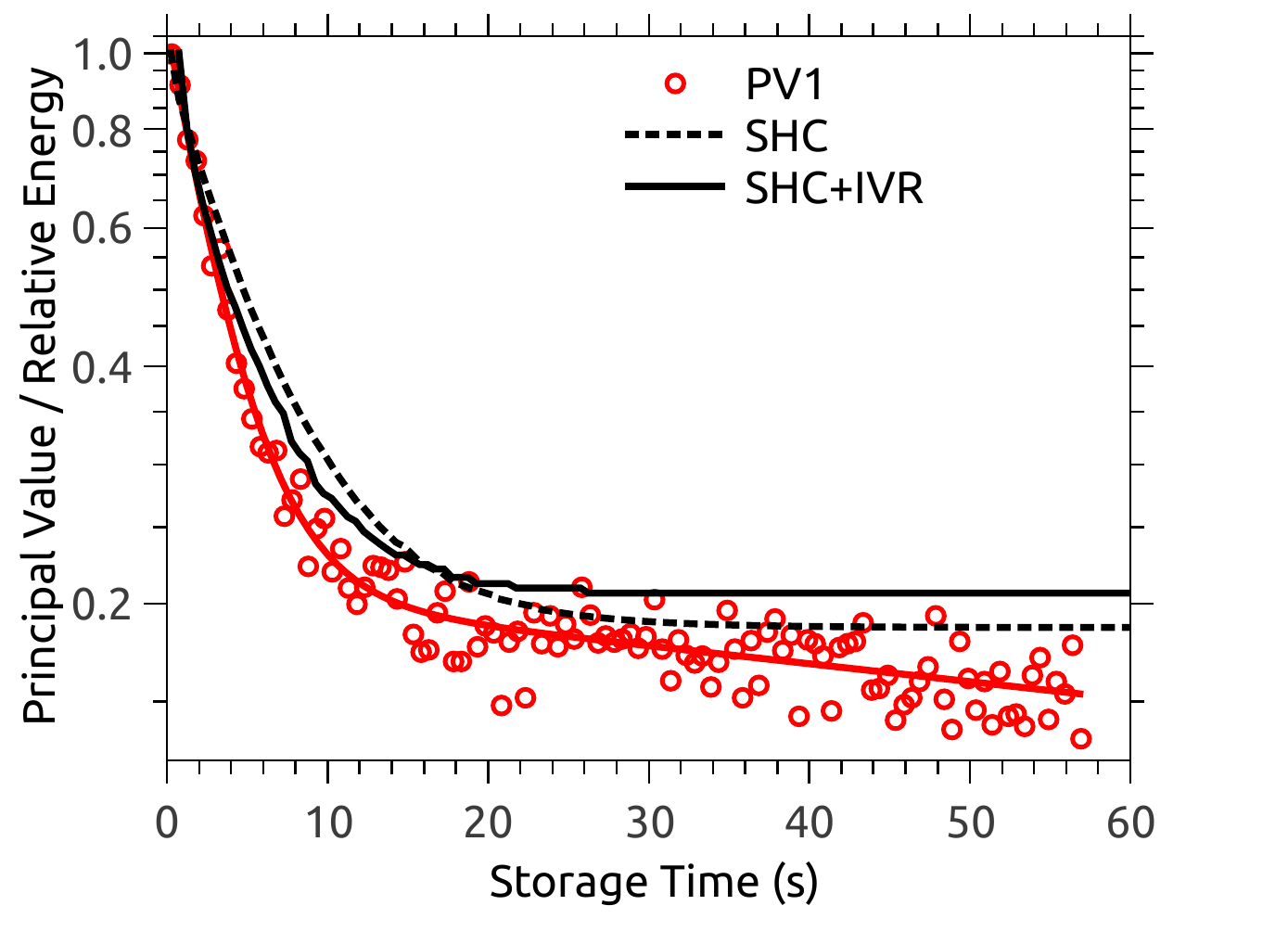}
  \caption{Principal component analysis on C$_{3}^{-}$. Upper: Principal component (PC1) and cold photodetachment spectrum. Lower: Principal values for PC1 (denoted PV1) with ion storage time, and simple harmonic cascade (SHC) model of the cooling lifetime (note the log scale). The gradual decrease in PV1 for ion storage times longer than $\approx$20\,s is attributed to the beam storage lifetime in DESIREE.}
  \label{fig_c3mod}
\end{figure}

Principal component analysis (PCA) of the 2D photodetachment spectrum for C$_3^-$ suggested a single principal component (PC1 in FIG.\,\ref{fig_c3mod}) can describe the hot-band intensity with ion storage time. Nearly 80$\%$ of the variance in the 2D spectrum is explained by PC1, with the remaining PCs describe only statistical fluctuations with no secular time dependence. The principal values of PC1 (denoted PV1) with ion storage time are shown in the lower panel of FIG.\,\ref{fig_c3mod}. Fit of PV1 with a bi-exponential gave a fast lifetime of 3.1$\pm$0.1\,s, which, as expected, is intermediate between the lifetimes for the wavelength-selected cooling times in FIG.\,\ref{fig_c3raw}. The second lifetime ($>$200\,s) is much longer than the measurement cycle (60\,s for C$_{3}^{-}$) and is presumably associated with the beam storage lifetime (540$\pm$30\,s for C$_3^-$, see Supporting Information). The time-invariant cold spectrum (FIG.\,\ref{fig_c3mod}, lower) was obtained by subtracting PC1, weighted by PV1, from the 2D photodetachment spectrum. This closely resembles the cold spectrum in FIG.\,\ref{fig_c3raw}, but utilizes the entire data set rather than arbitrarily time-binned data. Fit of the cold spectrum from PCA with the Wigner threshold law gave an ADE of 1.987$\pm$0.004\,eV (624.0$\pm$0.6\,nm), which is within error of above determination using the cold, time-binned photodetachment spectrum.

IR radiative cooling characteristics for C$_3^-$ from the SHC model are summarized in FIG.\,\ref{fig_c3mod}, lower. The dashed black curve assumes the case of no IVR and the solid black curve includes IVR. Exponential fits to the SHC curves returned ion cooling lifetimes of 5.22$\pm$0.01~s (no IVR) and 3.68$\pm$0.06~s (including IVR). The latter is in reasonable agreement with the average ion cooling lifetime from PCA (3.1$\pm$0.1\,s).

\begin{figure}[!t]
\centering
  \includegraphics[width=0.99\columnwidth]{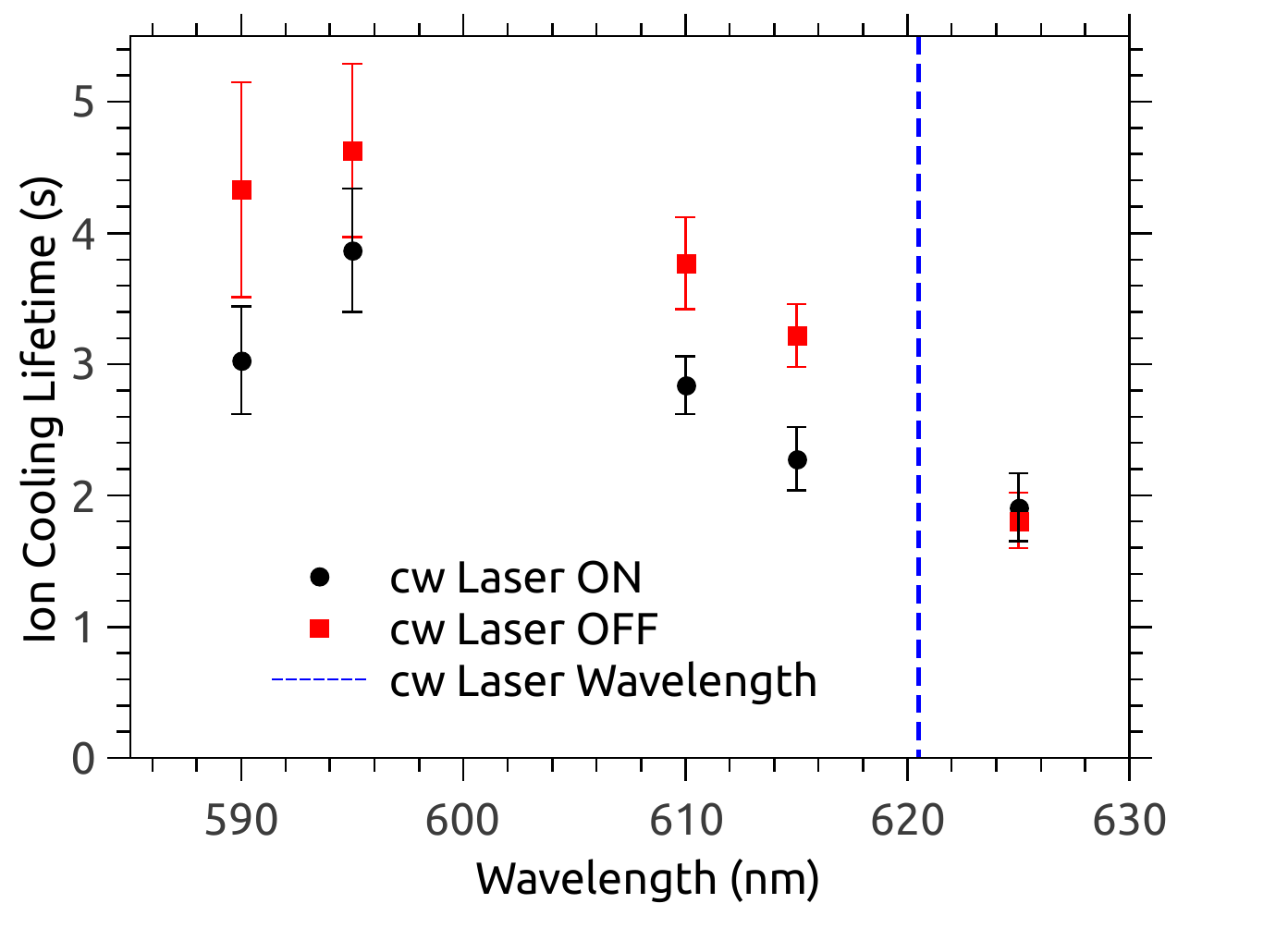}
  \caption{Cooling lifetimes for C$_3^-$ using the deplete-probe scheme. The black circles and red squares are cooling lifetimes with and without irradiation using cw laser light at 620.5~nm (dashed blue vertical line).}
  \label{fig_dye}
\end{figure}

The influence of the cw laser (620.5\,nm) on the cooling lifetimes of C$_3^-$ is shown in FIG.\,\ref{fig_dye}. Comparison of cw laser ON (black) with cw laser OFF (red) data at the probe wavelengths of 615, 610, 595 and 590\,nm show a systematic decrease of the ion cooling lifetimes by $\approx$1\,s because the photodetachment cross-section is larger for vibrationally excited ions than for cold ions at 620.5\,nm. No such effect was observed at 625\,nm, i.e. probe wavelength longer than that of the cw laser. These data provide a proof-of-principle measurement demonstrating a deplete-probe scheme to preferentially remove hot ions from the stored ion beam. The extent of depletion could likely be improved through better overlap of the cw beam with the ion beam and increase of cw laser power.

\subsection{Tetracarbon anion, C$_4^-$}

\begin{figure}[!t]
\centering
  \includegraphics[width=0.99\columnwidth]{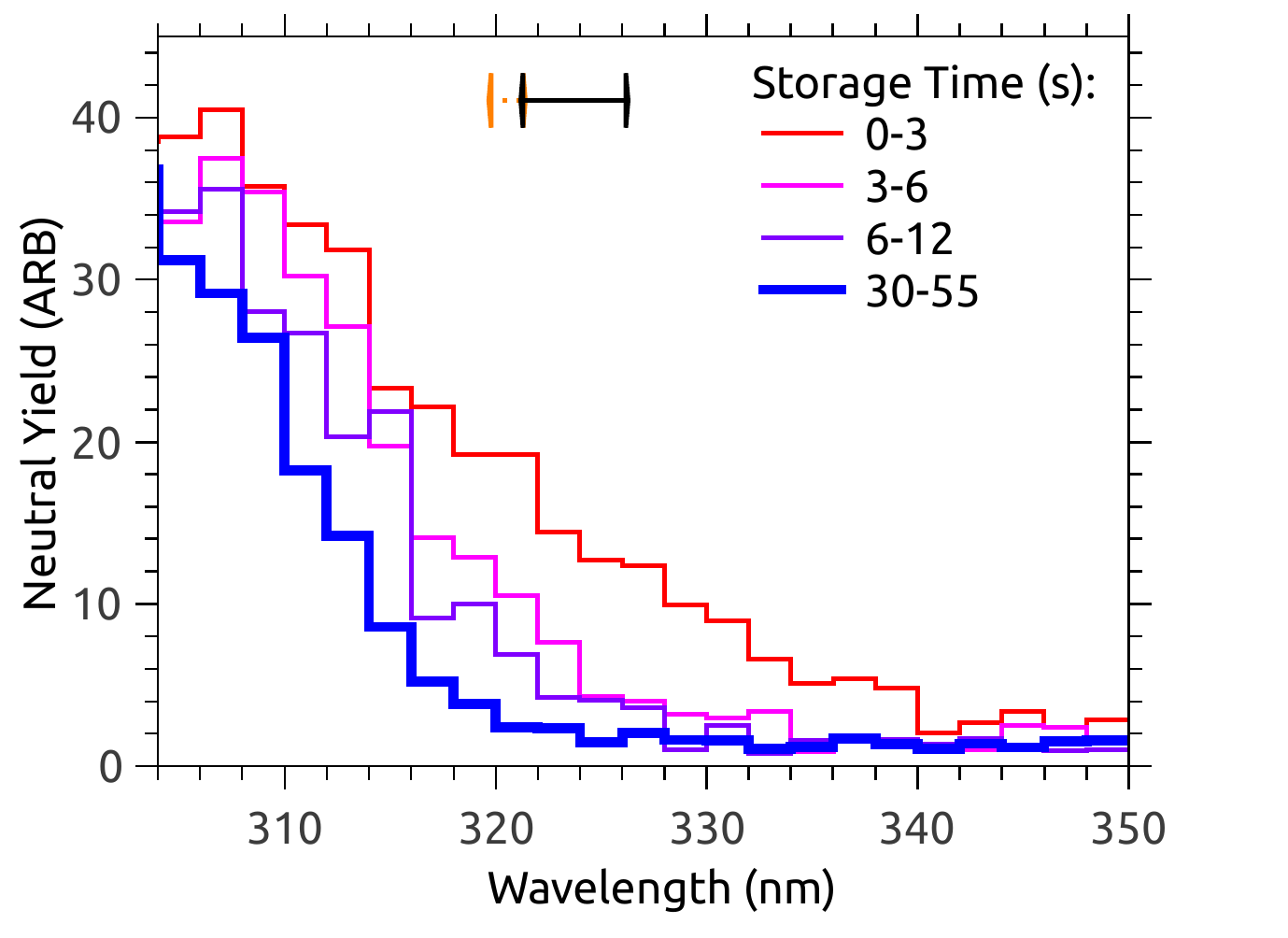}
  \includegraphics[width=0.99\columnwidth]{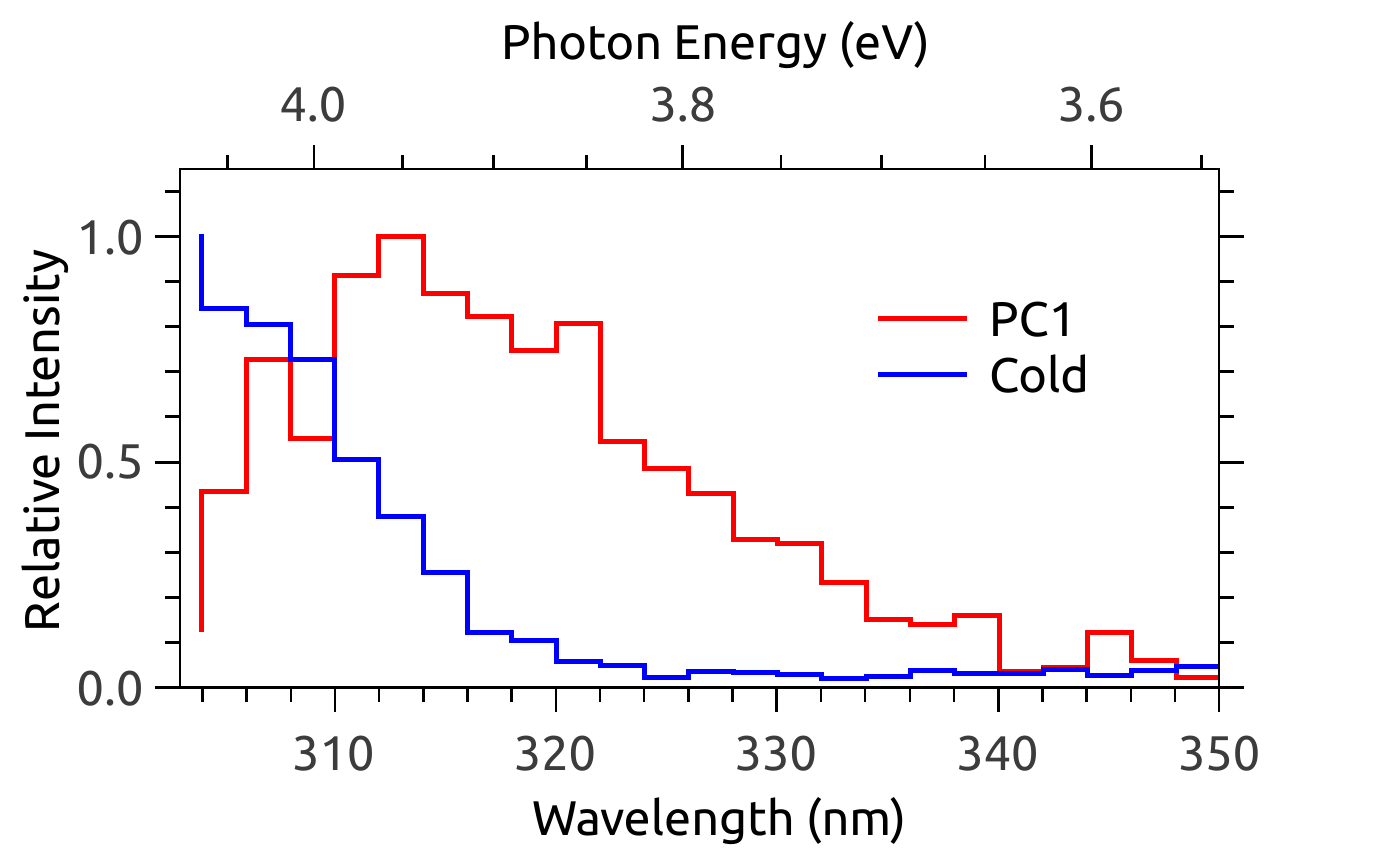}
  \includegraphics[width=0.99\columnwidth]{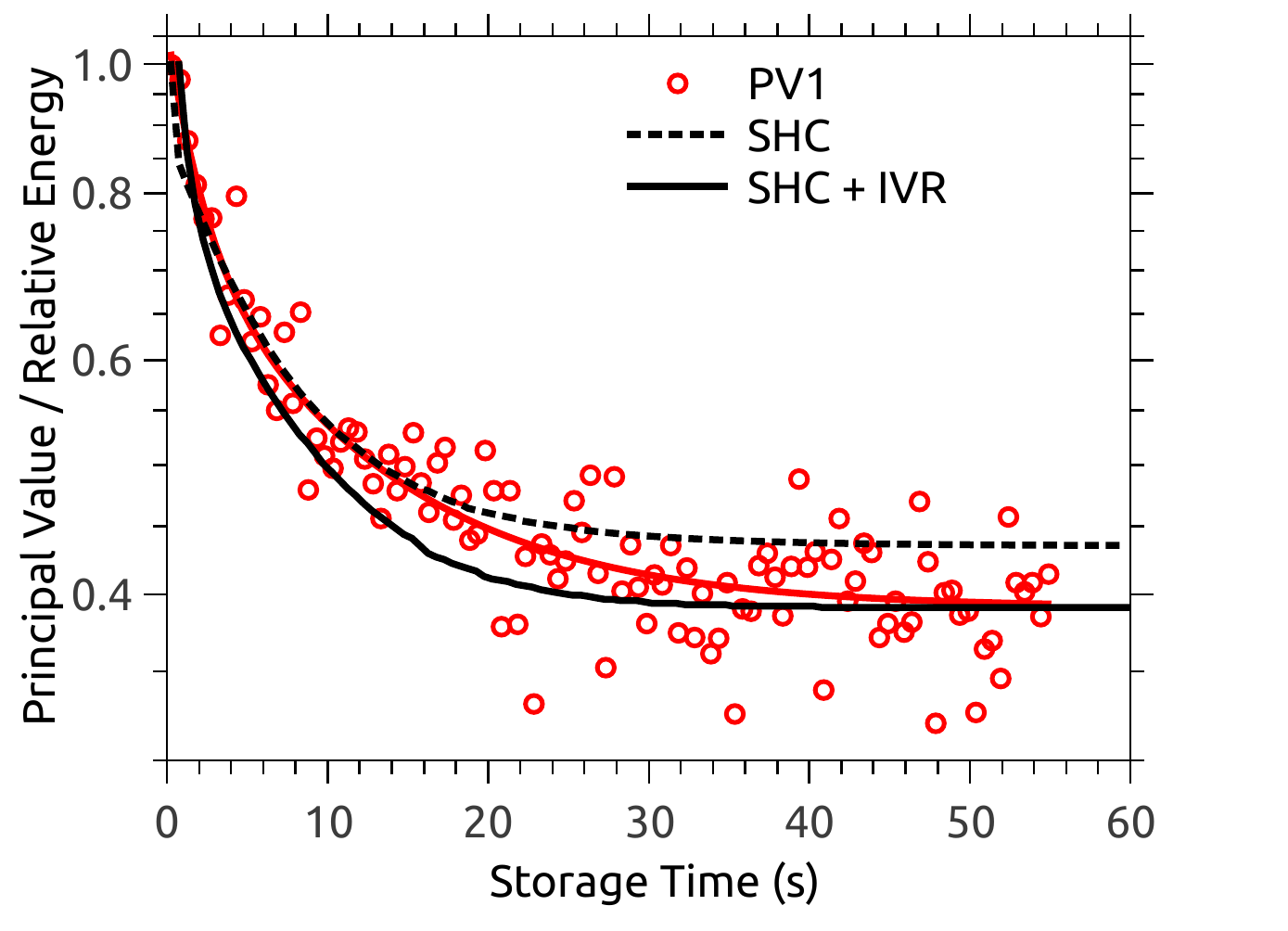}
  \caption{Upper: Time-binned photodetachment spectra for C$_4^-$, recorded by monitoring the yield of neutral particles with wavelength of light. The black bar represents the ADE and uncertainty determined for ions stored at least 30~s, and the orange bar corresponds to the photoelectron spectroscopy value from Ref.\citenum{Arnold:1991}. Middle: Principal component (PC1) and cold photodetachment spectrum from the 2D photodetachment spectrum of C$_4^-$. Bottom: Principal values for PC1 (denoted PV1) and simple harmonic cascade (SHC) model of the cooling lifetime (note the log scale).}
  \label{fig_c4cool}
\end{figure}

Time-binned photodetachment spectra and results from PCA for C$_4^-$ are summarized in FIG.\,\ref{fig_c4cool} upper and middle/lower, respectively. The C$_4^-$ photodetachment data were recorded in larger wavelength increments compared with C$_3^-$ or C$_5^-$ due to substantially lower laser fluence from the OPO at the near-UV wavelengths needed to photodetach this species. The time-binned photodetachment spectra for C$_{4}^{-}$ indicate that hot-band signal has disappeared after $\approx$30\,s. Fit of the 30--55\,s time-binned spectrum with the Wigner threshold law gave an ADE of 3.83$\pm$0.03\,eV (323.7$\pm$2.5\,nm), which is consistent with the value from photoelectron spectroscopy (3.882$\pm$0.010\,eV).\cite{Arnold:1991}


Application of PCA to the 2D photodetachment spectrum of C$_4^-$ again suggested that a single principal component (PC1 in FIG.\,\ref{fig_c4cool}, middle) describes the variation in the hot band intensity with ion storage time. The principal value of PC1 with ion storage time (PV1 in FIG.\,\ref{fig_c4cool}, lower) has a fitted lifetime of 6.8$\pm$0.5~s, which is roughly twice that for C$_{3}^{-}$ and comparable with the wavelength-binned values given above. Unfortunately, the data is of insufficient quality for a bi-exponential fit to account for the beam storage lifetime. As for C$_{3}^{-}$, shorter wavelengths are associated with longer ion cooling lifetimes -- see Supporting Information for further details.

IR radiative cooling lifetimes for C$_4^-$ from the SHC model are 6.74$\pm$0.01~s (no IVR) and 5.4$\pm$0.1~s (including IVR), with both models being in reasonable agreement with the PCA value of 6.8$\pm$0.5~s.

\subsection{Pentacarbon anion, C$_5^-$}

\begin{figure}[!t]
\centering
  \includegraphics[width=0.99\columnwidth]{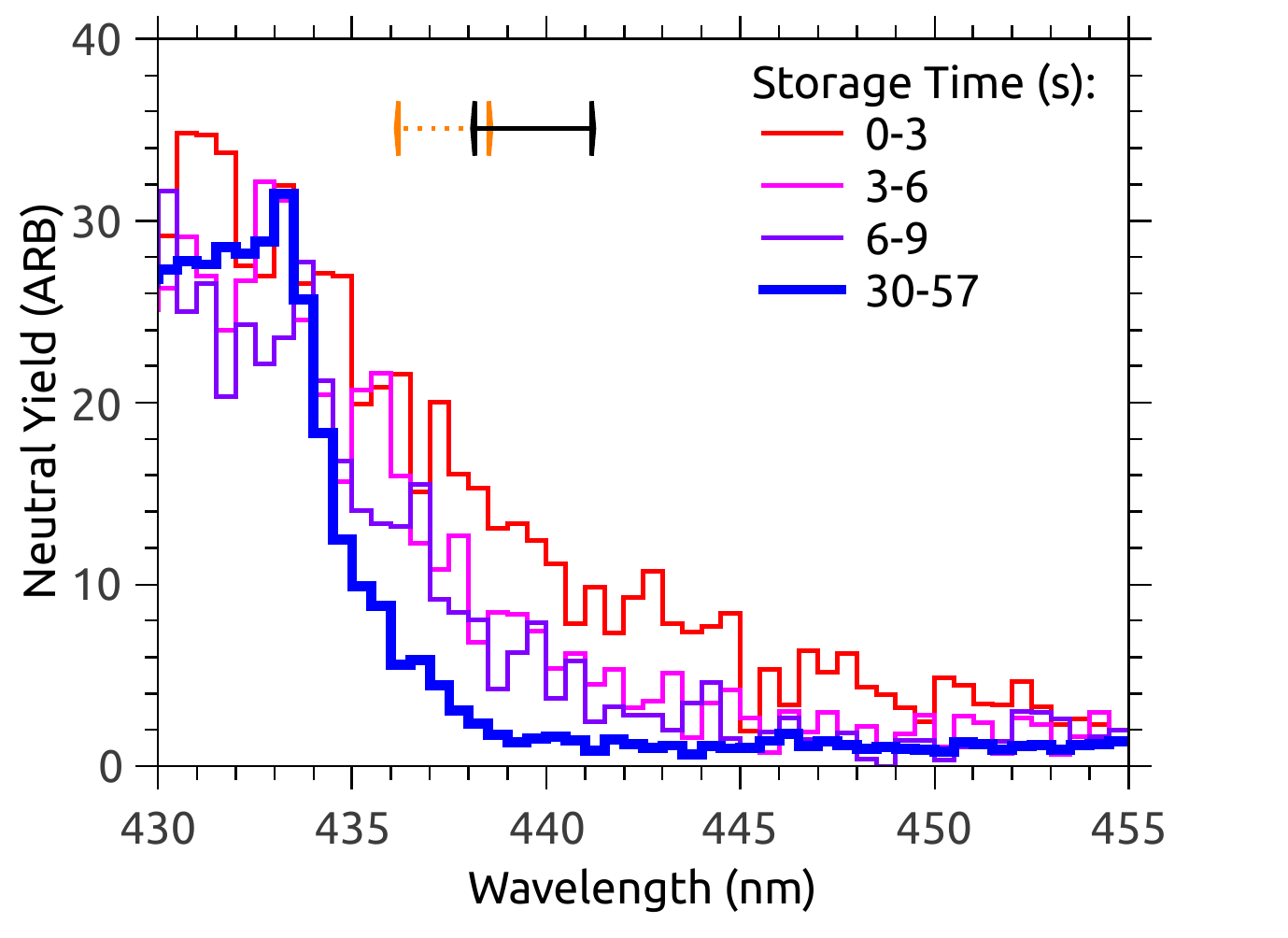}
  \includegraphics[width=0.99\columnwidth]{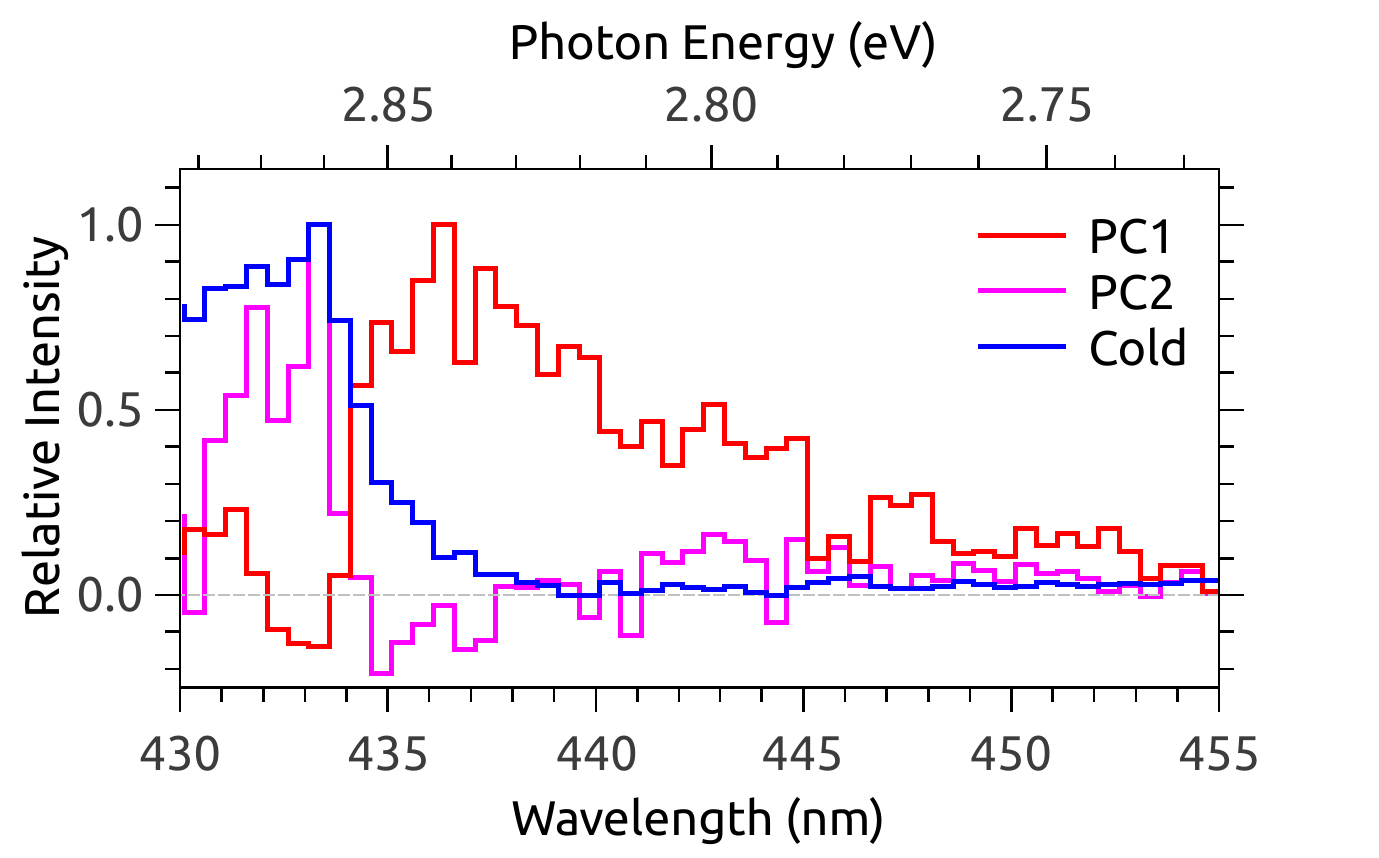}
  \includegraphics[width=0.99\columnwidth]{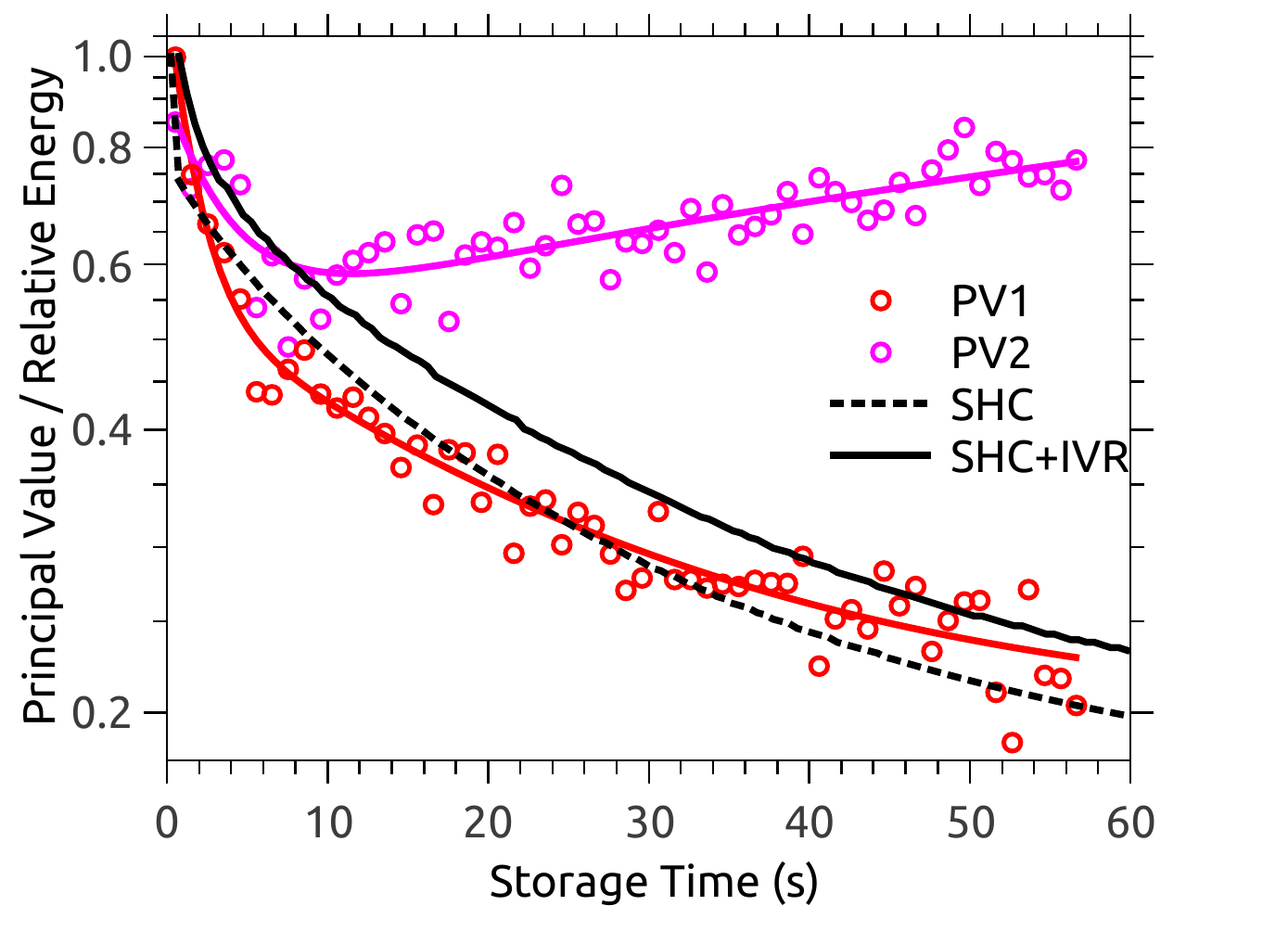}
  \caption{Upper: Time-binned photodetachment spectra for C$_5^-$, recorded by monitoring the yield of neutral particles with wavelength of light. The black bar represents the ADE and uncertainty determined for ions stored at least 30~s, and the orange bar corresponds to the photoelectron spectroscopy value from Ref.\citenum{Arnold:1991}. Middle: Principal components (PC1 and PC2) and cold photodetachment spectrum extracted from the 2D photodetachment spectrum of C$_5^-$.
Bottom: Principal values for PC1 and PC2 (denoted PV1 and PV2) and simple harmonic cascade (SHC) model of the cooling lifetime (note the log scale).}
  \label{fig_c5cool}
\end{figure}

Time-binned photodetachment spectra and PCA results for C$_5^-$ are summarized in FIG.\,\ref{fig_c5cool} upper and middle/lower, respectively. Fit of the 30--57\,s `cold' time-binned spectrum with the Wigner threshold law gave ADE = 2.82$\pm$0.01\,eV (439.7$\pm$1.6\,nm), which agrees with the value from photoelectron spectroscopy (2.839$\pm$0.008\,eV).\cite{Arnold:1991} Intriguingly, the cooling behaviour presents a different situation compared with C$_3^-$ and C$_4^-$. Whereas hot band photodetachment signal at wavelengths longer than $\approx$435\,nm diminishes over the first few seconds of ion storage, there is an enhancement of photodetachment signal for wavelengths shorter than $\approx$435\,nm (i.e. above the ADE), which will be discussed soon. 

Ion cooling lifetimes at selected probe wavelengths are 22$\pm$3\,s (435\,nm), 7.7$\pm$0.8\,s (440\,nm), and 5.7$\pm$0.6\,s (445\,nm) -- see Supporting Information for further details. As for C$_{3}^{-}$ and C$_{4}^{-}$, shorter wavelengths are associated with longer ion cooling lifetimes.

Application of PCA to the 2D photodetachment spectrum of C$_5^-$ suggested that two principal components (PC1 and PC2 in FIG.\,\ref{fig_c5cool}, middle) are necessary to describe the spectral variation with ion storage time. PC1 has a similar wavelength dependence and also principal value (PV1) with ion storage time when compared with PC1 for C$_3^-$ or C$_4^-$. PV1 was best fit with two exponential lifetimes of 1.7$\pm$0.3\,s and 24$\pm$5\,s, although there is also an unresolved long-lifetime decay associated with the beam storage lifetime. PC2 (FIG.\,\ref{fig_c5cool} middle) resembles a vibrational-like peak for wavelengths just shorter than the ADE wavelength. Fit of the principal values for PC2 with ion storage time (PV2 in FIG.\,\ref{fig_c5cool}, lower) required both exponential decay and growth functions. The growth lifetime for PC2 is within error of the long decay lifetime associated with PC1 (i.e., after $\approx$10\,s PV1 + PV2 is roughly steady state), implying that hot band population associated with PC1 eventually contributes to PC2 at longer ion storage time. We assign PC2 to predominately the $^{1}\Sigma_{g}^{+}(\nu^{\prime} = 0)\leftarrow {^{2}\Pi}_{\frac{3}{2}}(\nu^{\prime\prime} = 0)$ detaching transition, which occurs at slightly shorter wavelength ($\approx$0.5\,nm) than the alternative $^{1}\Sigma_{g}^{+}(\nu^{\prime} = 0)\leftarrow {^{2}\Pi}_{\frac{1}{2}}(\nu^{\prime\prime} = 0)$ spin-orbit detaching transition.\cite{Arnold:1991,Kitsopoulos:1991,Hock:2012} Assuming this assignment is correct, it appears that as ions cool, the relative population of ground vibrational state anions increases and consequently the apparent photodetachment cross-section for resonant detaching transitions increases (presumably much more so than for C$_{3}^{-}$ and C$_{4}^{-}$). It follows that the long lifetime associated with PC2 is due to decay of population associated with the $^{1}\Sigma_{g}^{+}(\nu^{\prime}_{7} = 1)\leftarrow {^{2}\Pi}_{\frac{3}{2}}(\nu^{\prime\prime}_{7} = 1)$ and $^{1}\Sigma_{g}^{+}(\nu^{\prime}_{7} = 1)\leftarrow {^{2}\Pi}_{\frac{1}{2}}(\nu^{\prime\prime}_{7} = 1)$ hot band detaching transitions (see spectroscopic assignment of photodetaching hot band modes in Refs.\,\citenum{Arnold:1991,Kitsopoulos:1991,Hock:2012}). Note, because $\nu^{\prime\prime}_{7}$ is an IR inactive mode (see Supporting Information), decay must be due to IVR followed by radiative emission.

IR radiative cooling characteristics for C$_5^-$ from the SHC model are summarized in FIG.\,\ref{fig_c5cool}, lower. Neglect of IVR resulted in a cooling curve that was best fit with two lifetimes of 4.78$\pm$0.08~s and 21.7$\pm$0.1~s. Inclusion of IVR gave fitted lifetimes of 1.63$\pm$0.04~s and 22.0$\pm$0.2~s, which are in good agreement with values from PCA. The need for a bi-exponential fit for C$_{5}^{-}$ can be traced to mode-specific radiative emission processes. Specifically, the faster lifetime is dominated by emission from the main IR active mode $\nu_{9} \approx 1751$\,cm$^{-1}$ ($A_{9}^{10}\approx 1807$) and the slower lifetime attributed to emission from weaker modes $\nu_{1,2} \approx 127-141$\,cm$^{-1}$ ($A_{1,2}^{10}\approx 21-30$) -- see mode-specific radiated power plots in the Supporting Information. Similar double lifetime cooling is not apparent for C$_3^-$ and C$_4^-$ because the majority of the cooling from the high frequency mode with a large $A^{10}$ coefficient occurs on a sub-second timescale (see Supporting Information).

\section{Summary and outlook}

The present work has investigated the ultraslow cooling characteristics of three astrochemically relevant anions under conditions approximating a molecular cloud. Interestingly, an increase in molecular size leads to longer average ion cooling lifetimes: 3.1$\pm$0.1\,s for C$_3^-$, 6.8$\pm$0.5\,s for C$_4^-$ and 24$\pm$5\,s for C$_5^-$. Variation in ion cooling lifetimes across the hot band is attributed to a distribution of internal energies. These are the first known measurements on carbonaceous anions extending to the ultraslow (seconds) timescale; all previous measurements have been performed under room temperature conditions and were restricted to measuring the sub-second cooling dynamics.

The increase in average ion cooling lifetime with molecular size can be understood by considering the point group symmetry (D$_{\infty h}$) of the anions and that E1 radiative transitions require a change in electric dipole moment. In particular, the high symmetry means that each of present anions have only three vibrational modes with $A^{10}$ coefficents larger than 10. Although $A^{10}$ coefficients for $^{1}\Sigma_{g}^{-}$ symmetric vibrational modes quickly increase with molecular size beyond C$_{5}^{-}$, for $n=3-5$ there are an increasing number of low frequency modes with increasing $n$ (mostly IR inactive) and the weakly IR active $\Pi_{u}$ symmetric vibrational modes become lower in frequency and have lower radiative emission rates (see EQN\,2 and Supporting Information). The net result is an increase in ion cooling lifetime with increasing $n$. These ion cooling dynamics would not be evident at room temperature for $n>3$ because the average thermal vibrational energy exceeds the energy of the low frequency modes: $\approx$342\,cm$^{-1}$ (C$_{4}^{-}$) and $\approx$587\,cm$^{-1}$ (C$_{5}^{-}$) at 298\,K assuming harmonic vibrational partition functions. We are presently applying the 2D photodetachment strategy with DESIREE to study the ultraslow cooling dynamics of larger carbonaceous anions to further explore these trends. These results will be the presented in a forthcoming paper.

As part of the present study, we developed a simple harmonic cascade model that proved capable of simulating IR radiative emission using input data from conventional electronic structure calculations. With provision for IVR, the model was able to qualitatively reproduce the experimental ion cooling lifetimes and provide a mode-by-mode understanding of the cooling dynamics. The agreement between theory and experiment provides confidence for applying this model to anions for which experimental data is not available or difficult to measure. 

Finally, it should be noted that application of the present 2D photodetachment methodology to larger molecular anions may prove more complicated due to near-threshold resonant excitations. Specifically, if there are substantial cross-sections for photoexcitation of $\pi\pi^{*}$ states situated below the detachment threshold or for resonances situated in the detachment continuum, ensuing autodetachment and internal conversion dynamics might affect the observed ion cooling lifetimes and spectral features. For example, experiments have shown that photoexcitation followed by internal conversion to recover the ground electronic state is efficient for C$_n^{-}$ ($n>4$) and polycyclic aromatic hydrocarbon (PAH) anions,\cite{Zhao:1996,Baguenard:2002,menadione,tetracene} with photoexcitation cross-sections for optically allowed transitions in PAH anions generally being much larger than cross-sections for direct photodetachment. If neutral formation through thermionic emission or dissociation processes takes longer than the time ions spend in the straight section of the ion storage ring after irradiation ($\approx$4--5\,$\mu$s), then neutrals formed outside of the straight section of the ion storage ring will not be counted. Fortunately, in DESIREE, the relative importance of delayed neutral formation can be ascertained by simultaneously measuring neutral yield on the detector on the opposite straight section of the ion storage ring (Glass Plate/MCP detector in FIG.\,\ref{fig:desiree}).

\section*{Acknowledgements}
This work was supported by the Swedish Research Council (grant numbers 2015-04990, 2016-03675, 2016-04181, 2018-04092) and the Swedish Foundation for International Collaboration in Research and Higher Education (STINT, grant number PT2017-7328 awarded to JNB, EC and MHS). We acknowledge the DESIREE infrastructure for provisioning of facilities and experimental support, and thank the operators and technical staff for their invaluable assistance. The DESIREE infrastructure receives funding from the Swedish Research Council under the grant number 2017-00621.

\bibliography{cnm}

\end{document}